\documentstyle[art11]{article}
\newcommand{\GRACE}{{\tt GRACE}\ }
\newcommand{\REDUCE}{{\tt REDUCE}}
\newcommand{\CHANEL}{{\tt CHANEL}}
\newcommand{\BASES}{{\tt BASES/SPRING}\ }

\def\selecR{\tilde e_{\rm R}}
\def\selecL{\tilde e_{\rm L}}

\def\lsp{{\tilde\chi^0}_1}
\def\cinolm{\tilde{\chi_{\rm 1}}^-}
\def\cinolp{\tilde{\chi_{\rm 1}}^+}
\def\photino{\tilde\gamma}
 
\begin{document}
\renewcommand{\thefootnote}{*)}
 
\begin{center}
{\Large \bf Automatic Calculation of SUSY-particle
Production}\footnote{Talk presented by M.Jimbo at
Xth International Workshop for HEP and QFT, Zvenigorod,
September 1995.}
\\

\vspace*{2cm}

\font\fonts=cmbx10
{\large \bf Masato J{\fonts IMBO}}\\
{\it Computer Science Laboratory, Tokyo Management College,
625-1 Futamata}\\
{\it Ichikawa, Chiba 272, JAPAN}\\
{\it (e-mail: jimbo@tmc-ipd.ac.jp)}\\
\vspace*{11pt}
{\large \bf Hidekazu T{\fonts ANAKA}}\\
{\it Faculty of Science, Rikkyo University, 3-34-1 
Nishi-ikebukuro}\\
{\it Toshima, Tokyo 171, JAPAN}\\
{\it (e-mail: tanakah@minami.kek.jp)}\\
\vspace*{11pt}
{\large \bf Tadashi K{\fonts ON}}\\
{\it Faculty of Engineering, Seikei University, 3-3-1
Kichijoji-kita}\\
{\it Musashino, Tokyo 180, JAPAN}\\
{\it (e-mail: kon@ge.seikei.ac.jp)}\\
\vspace*{11pt}
{\large \bf Toshiaki K{\fonts ANEKO}}\\
{\it Faculty of General Education, Meiji-gakuin University,
1518 Kami-kurata}\\
{\it Totsuka, Yokohama 241, JAPAN}\\
{\it (e-mail: kaneko@minami.kek.jp)}\\
\vspace*{11pt}
{\large \bf M{\fonts INAMI}-T{\fonts ATEYA} collaboration}\\
\end{center}

\vspace*{1cm}

\begin{abstract}
 We introduce a new method to treat Majorana fermions
and interactions with fermion-number violation on the
\GRACE system which has been developed for the automatic
computation of the matrix elements for the processes of
the standard model.  Thus we have constructed a system for
the automatic computation of cross-sections for the
processes of the minimal SUSY standard model (MSSM).
\end{abstract}
\renewcommand{\thefootnote}{\sharp\arabic{footnote}}
\renewcommand{\theequation}{\arabic{section}.\arabic{equation}}
\setcounter{footnote}{0}
 
\section{Introduction}
 
 At the theoretical point of view, it has been a promising hypothesis that
there exists a symmetry called supersymmetry (SUSY) between bosons and
fermions at the unification-energy scale.  It, however, is a broken symmetry
at the electroweak-energy scale.  Thus the relic of SUSY is expected to remain
as a rich spectrum of SUSY particles, partners of usual matter fermions, gauge
bosons and Higgs scalars, named sfermions, gauginos and higgsinos,
respectively~\cite{theor}.

 The quest of these new particles has already been one of the most important
pursuits to the present high-energy physics~\cite{exp}.  Although such
particles have not yet been discovered, masses of them are expected to be
$O(10^2)$ GeV~\cite{tub}.  In order to obtain signatures of the SUSY-particle
production, electron-positron colliding experiments are preferable because the
electroweak interactions are clean and well-known.  Thus we hope SUSY particles
will be detected at future $e^-e^+$-colliders of sub-TeV-region or TeV-region
energies such as LEP2 or NLC's (Next Linear Colliders)~\cite{col,Rev}.

 For the simulations of the experiments, we have to calculate the
cross-sections for the processes with the final 3-body or more.  We have
already known within the standard model that the calculation of the helicity
amplitudes is more advantageous to such a case than that of the traces for the
gamma matrices with \REDUCE~\cite{tks,wmh}.  The program package
\CHANEL~\cite{chan} is one of the utilities for the numerical calculation of
the helicity amplitudes, which has been developing by one of the authors
(H.T.).

 It, however, is also hard work to construct a program with many subroutine
calls of \CHANEL~by hand.  Thus we need a more convenient way to carry out such
a work.  Several groups have started independently to develop computer systems
which automate the perturbative calculation in the standard model with
different methods~\cite{pre-grc,comp,Den,SL,Den2}.  The \GRACE
system~\cite{pre-grc}, which automatically generates the source code for
\CHANEL, is one of the solutions.  The system also includes the interface and
the library of \CHANEL, and the program package \BASES v5.1~\cite{bas} for
multi-dimensional integrations and event-generations.

 In the SUSY models, there exist Majorana fermions as the neutral gauginos and
higgsinos, which become the mixed states called neutralinos.  Since
anti-particles of Majorana fermions are themselves, there exists so-called
`Majorana-flip', the transition between particle and anti-particle.  This is
the most important problem which we should solve when we realize the
automatic system for computation of the SUSY processes.

 In a recent work~\cite{jt,jtkk}, we developed an algorithm to treat
Majorana fermions in \CHANEL.  In the standard model, we already have such
particles as Dirac fermions, gauge bosons and scalar bosons in the \GRACE
system.  There, however, exists another problem on fermion-number violating
interactions.  We have also developed an algorithm for this problem.  Thus we
can construct an automatic system for the computation of the SUSY processes by
the algorithms above in the \GRACE system.

\section{SUSY particles and interactions into GRACE~2.0}

 In Fig.~1, we present the system flow of \GRACE (after version 1.1)~\cite{gm}.
The \GRACE system has become more flexible for the extension in the new version
called `{\tt grc}'~\cite{grc-pp}, which is written by {\tt C}, and includes a
new graph-generation package~\cite{Fdg}.  With this package, any graphs based
on a user-defined model can be generated at any orders.  The Feynman diagrams
are drawn by the program package `{\tt gracefig}'~\cite{sk} in the new
{\tt GRACE}.  It is necessary for us to make the interface and the library of
\CHANEL~ and the model file for including the SUSY particles.

\begin{figure}[htp]
\setlength{\unitlength}{1mm}
\begin{picture}(150,180)

\put(80,170){\framebox(50,8)[c]{\hbox{Theory(Lagrangian)}}}
\put(10,170){\framebox(50,8)[c]{\hbox{User input}}}
\put(10,155){\framebox(50,15)[c]{\hbox{Process(particle, order)}}}
\put(80,155){\framebox(50,8)[c]{\hbox{Particles and interactions}}}

\put(105,170){\vector(0,-1){7}}
\put(105,155){\vector(0,-1){5}}
\put(35,155){\vector(0,-1){5}}
\put(20,155){\vector(0,-1){40}}

\put(31,141){\framebox(78,8)[c]{\  }}
\put(30,140){\framebox(80,10)[c]{\hbox{Diagram generator}}}

\put(70,140){\vector(0,-1){7}}
\put(70,125){\vector(0,-1){25}}

  \put(50,125){\framebox(40,8)[c]{\hbox{Diagram description}}}
  \put(101,126){\framebox(38,8)[c]{ }}
  \put(100,125){\framebox(40,10)[c]{\hbox{Drawer}}}
  \put(100,105){\dashbox(40,10)[c]{\hbox{Feynman diagrams}}}
    \put(90,130){\vector(1,0){10}}
    \put(105,125){\vector(0,-1){10}}

  \put(5,105){\framebox(40,10)[c]{}}
   \put(6,108){\hbox{Kinematics database}}
  \put(23,105){\vector(0,-1){26}}

\put(31,91){\framebox(78,8)[c]{\  }}
\put(30,90){\framebox(80,10)[c]{\hbox{Matrix element generator}}}

  \put(15,65){\framebox(120,20)[c]{\  }}
    \put(20,67){\framebox(25,12)[c]{\  }}
    \put(54,67){\framebox(27,12)[c]{\  }}
    \put(90,67){\framebox(25,12)[c]{\  }}
    \put(28,80){\hbox{Function for matrix element}}
    \put(80,80){\hbox{\footnotesize (FORTRAN)}}
    \put(22,69){\hbox{code}}
    \put(22,73){\hbox{kinematics}}
    \put(57,69){\hbox{code}}
    \put(57,73){\hbox{generated}}
    \put(92,69){\hbox{library}}
    \put(92,73){\hbox{CHANEL}}
    \put(45,73){\line(1,0){9}}
    \put(90,73){\line(-1,0){9}}
    
\put(70,90){\vector(0,-1){11}}
\put(70,65){\vector(0,-1){5}}

\put(41,51){\framebox(78,8)[c]{\  }}
\put(40,50){\framebox(80,10)[c]{\hbox{BASES(Monte-Carlo integral)}}}

  \put(5,50){\framebox(23,10){\  }}
    \put(6,51){\hbox{information}}
    \put(6,55){\hbox{convergence}}
  \put(40,55){\vector(-1,0){12}}

  \put(5,28){\dashbox(45,15){\  }}
    \put(15,32){\hbox{Distributions}}
    \put(15,37){\hbox{Cross section}}
  \put(45,50){\vector(0,-1){7}}

  \put(55,50){\vector(0,-1){26}}
  \put(20,9){\framebox(40,15){\  }}
    \put(22,13){\hbox{Parameters}}
    \put(22,18){\hbox{Distribution}}
  \put(60,22){\vector(1,0){10}}

\put(71,21){\framebox(78,8)[c]{\  }}
\put(70,20){\framebox(80,10)[c]{\hbox{SPRING(event generator)}}}

\put(130,65){\vector(0,-1){35}}
\put(110,20){\vector(0,-1){5}}

  \put(85,5){\dashbox(50,10){\  }}
    \put(95,6){\hbox{generated events}}
    \put(95,10){\hbox{Specified number of}}
\end{picture}
  \hspace*{2cm} Fig.~1. GRACE system flow (after version 1.1)
\end{figure}

 The method of computation in the program package \CHANEL~ is as follows:
\begin{enumerate}
  \item To divide a helicity amplitude into vertex amplitudes.
  \item To calculate each vertex amplitude numerically as a complex number.
  \item To reconstruct of them with the polarization sum, and calculate
  the helicity amplitudes numerically.
\end{enumerate}
The merit of this method is that the extension of the package is easy,
and that each vertex can be defined only by the type of concerned particles.

 Here we use an algorithm~\cite{jt,jtkk} for the implementation of the
embedding Majorana fermions in \CHANEL~ as follows:
\begin{itemize}
  \item \underline{\bf policy}
  \begin{enumerate}
    \item To calculate a helicity amplitude numerically.
    \item To replace each propagator by wave functions or polarization vectors,
    and calculate vertex amplitudes.
    \item \underline{\bf Not to} move charge-conjugation matrices.
  \end{enumerate}
    \item \underline{\bf method}
  \begin{enumerate}
    \item To choose a direction on a fermion line.
    \item To put wave functions, vertices and propagators along the direction
    in such a way:
  \begin{itemize}
    \item[~i)] To take the transpose for the reverse direction of
    fermions
    \item[ii)] To use the propagator with the charge-conjugation
    matrix for\\
    the Majorana-flipped one.
  \end{itemize}
  \end{enumerate}
\end{itemize}

As a result, the kinds of the Dirac-Majorana-scalar vertices are limited to
four types:
\begin{eqnarray}
 J_{1~{h_1}{h_2}}^{[{\rm S}]{\rho_1}{\rho_2}} & = &
  \overline{U}^{\rho_1}({h_1},{p_1},{m_1}) \Gamma
  U^{\rho_2}({h_2},{p_2},{m_2})~~, \\
 J_{2~{h_1}{h_2}}^{[{\rm S}]{\rho_1}{\rho_2}} & = &
  U^{{\rho_1}\rm T}({h_1},{p_1},{m_1}) \Gamma
  \overline{U}^{{\rho_2}\rm T}({h_2},{p_2},{m_2})~~, \\
 J_{3~{h_1}{h_2}}^{[{\rm S}]{\rho_1}{\rho_2}} & = &
  \overline{U}^{\rho_1}({h_1},{p_1},{m_1}) C^{\rm T} \Gamma^{\rm T}
  \overline{U}^{{\rho_2}\rm T}({h_2},{p_2},{m_2})~~, \\
 J_{4~{h_1}{h_2}}^{[{\rm S}]{\rho_1}{\rho_2}} & = &
  U^{{\rho_1}\rm T}({h_1},{p_1},{m_1}) \Gamma^{\rm T} C^{-1}
  U^{\rho_2}({h_2},{p_2},{m_2})~~,
\end{eqnarray}
where $U$'s denote wave functions, and $C$ is the charge-conjugation matrix.
The symbol $\Gamma$ stands for the scalar vertex such as
\[ \Gamma = A_{\rm L}\cdot{{1 - \gamma_5}\over{2}} +  
A_{\rm R}\cdot{{1 + \gamma_5}\over{2}} ~~. \]
The vertices $J_{2}^{[{\rm S}]} \sim J_{4}^{[{\rm S}]}$ are related to the
vertex $J_{1}^{[{\rm S}]}$ which has been already defined as the
Dirac-Dirac-scalar vertex in the subroutine FFS of \CHANEL.  The relations
among the vertices are as follows:
\begin{eqnarray}
 J_{1~{h_1}{h_2}}^{[{\rm S}]{\rho_1}{\rho_2}} \qquad & & \rightarrow
 {\rm FFS}~~,\\
 J_{2~{h_1}{h_2}}^{[{\rm S}]{\rho_1}{\rho_2}} \quad = &
  -J_{1~{h_1}{h_2}}^{[{\rm S}]{-\rho_1}{-\rho_2}}~ & \rightarrow
  {\rm FFST}~~,\\
 J_{3~{h_1}{h_2}}^{[{\rm S}]{\rho_1}{\rho_2}} \quad = &
  -J_{1~{h_1}{h_2}}^{[{\rm S}]{\rho_1}{-\rho_2}} \quad & \rightarrow
  {\rm FFCS}~~,\\
 J_{4~{h_1}{h_2}}^{[{\rm S}]{\rho_1}{\rho_2}} \quad = &
  -J_{1~{h_1}{h_2}}^{[{\rm S}]{-\rho_1}{\rho_2}} \quad & \rightarrow
  {\rm FFSC}~~,
\end{eqnarray}
Thus we can build three new subroutines {\tt FFST}, {\tt FFCS}~ and {\tt FFSC}.
We have performed the installation of the subroutines above with the interface
on the \GRACE system version 2.0~\cite{grc2.0b,jtkk,YITP,mj,Mor,ai95}.

 Next we propose an algorithm for the interactions with fermion-number
violation such as the chargino-selectron-antineutrino vertex.  We introduce
two new soubroutines~\cite{prep,Appi}.

\section{Tests for the system}

 At the first stage for the check of our system, we have written the model file
of the SUSY QED.  In this case, there is only one Majorana fermion called
photino.  It is essential for testing our system to include photino and its
interactions.

 Next we have extended the model file and the definition file of couplings for
the MSSM.  The tests have been performed by the exact calculations with the two
methods, our system and \REDUCE, in such a manner:
\begin{enumerate}
  \item To calculate the differential cross-sections at a point of the
  phase space in the two methods with \GRACE and \REDUCE.
  \item To calculate the differential cross-sections over the
  phase space in the two methods with \GRACE and \REDUCE~ through {\tt BASES}.
  \item To integrate the differential cross-sections over the
  phase space in the two methods with \GRACE and \REDUCE~ through {\tt BASES}.
\end{enumerate}
With {\tt BASES}, we can get the differential cross-sections and the scattered
plots by one time of the integration step.  For writing \REDUCE~ sources, we
use the different method to treat Majorana fermions in Ref. \cite{Den2}.

 In Table I, the tested processes are shown as a list.  The references in the
table (without \cite{jtkk}, \cite{Mor} and \cite{jlc5}) are not the results of
the tests, but for help.

\begin{table}[hbt]
\begin{center}
  \begin{tabular}{llclcc}  \hline
  Process &  & Number of diagrams & Comment & Check & Reference \\
  \hline\hline
  {\bf SUSY} & {\bf QED} & & & & \\ \hline
$e^- e^- \rightarrow$ & $\selecR^- \selecR^-$ & 2 & Majorana-flip & OK &
        \cite{jtkk}\\
        & $\selecL^- \selecL^-$ & 2 & in internal lines & OK & \cite{jtkk} \\
                   & $\selecR^- \selecL^-$ & 2 & & OK & \cite{jtkk}\\  \hline
$e^- e^+ \rightarrow$ & $\selecR^- \selecR^+$ & 2 & Including pair & OK &
		    \cite{mj,HESP}\\
     & $\selecL^- \selecL^+$ & 2 & annihilation & OK & \cite{mj,HESP} \\ \hline
$e^- e^+ \rightarrow$ & $\selecR^- \selecL^+$ & 1 & Values are & OK &
		    \cite{mj,HESP} \\
        & $\selecR^+ \selecL^-$ & 1 & equal & OK & \cite{mj,HESP} \\  \hline
$e^- e^+ \rightarrow$ & $\photino \photino$ & 4 & F-B symmetric & OK &
	\cite{jtkk} \\  \hline
$e^- e^+ \rightarrow$ & $\photino \photino \gamma~$ & 12 & Final 3-body
   & OK & \cite{tk} \\ \hline
$e^- e^+ \rightarrow$ & $\selecR^- \photino e^+$ & 12 & Including every
   & OK & \cite{spu,epa} \\
   & & & elements for tests & & \cite{jlc5} \\ \hline\hline
  {\bf MSSM} & & & & \\ \hline
$e^- e^- \rightarrow$ & $\selecL^- \selecL^-$ & 8 & 4 Majorana fermions & OK &
   \cite{Mor} \\ \hline
$e^- e^+ \rightarrow$ & $\cinolm \cinolp$ & 3 & & OK & \cite{Mor} \\  \hline
$e^- e^+ \rightarrow$ & $\lsp \lsp$ & 14 & Final 3-body & OK & \cite{Mor}
\\  \hline
  \end{tabular}
\end{center}
  \hspace*{4cm} Table~I. The list of the tested processes.
\end{table}%

 In Ref.~\cite{jlc5}, we show the angular distribution of the outgoing positron
in the process $e^- e^+ \rightarrow \selecR^- \photino e^+$.  Here we use
{\tt BASES}~ for the calculation from the \REDUCE~ output. The result is in
beautiful agreement with the value that is obtained by \GRACE at each bin of
the histogram.  Since the two diagrams with the one-photon exchange dominate in
this case, there is a steep peak in the direction of the initial positron.  In
such a case, the equivalent-photon approximation (EPA) works well~\cite{epa}.

\section{Summary}

 We introduce a new method to treat Majorana fermions and interactions with
fermion-number violation on the \GRACE system for the automatic computation of
the matrix elements for the processes of the SUSY models.  In the first
instance, we have constructed the system for the processes of the SUSY QED
because we should test our algorithm for the simplest case.  Next we have
extended the model file and the definition file of couplings for the MSSM.  The
numerical results convince us that our algorithm is correct.

\section{Acknowledgements}

  This work was supported in part by the Ministry of Education,
  Science and Culture, Japan under Grant-in-Aid for International
  Scientific Research Program No.04044158 and No.07044097.  Two of the
  authors (H.T. and M.J.) have been also indebted to the
  above-mentioned Ministry under Grant-in-Aid for Scientific Research (C)
  Program  No.06640411.


\begin{thebibliography}{99}
  \bibitem{theor} H.P. Nilles, {\sl Phys. Rep.} {\bf 110} (1984), 1.\\
   H.E. Haber and G.L. Kane, {\sl Phys. Rep.} {\bf 117} (1985), 75.\\
   M. Chen, C. Dionisi, M. Martinez and X. Tata, {\sl Phys. Rep.}
  {\bf 159} (1988), 201.\\
   R. Barbieri, {\sl Riv. Nuovo Cimento} {\bf 11} (1988).\\
   R. Barbieri {\it et al.}, {\sl Z PHYSICS AT LEP 1},
  CERN Report {\sl CERN} {\bf 89-08} Vol.{\bf 2} (1989), p.121.
  \bibitem{exp} C. Dionisi, in {\sl Proceedings of XVII International Meeting on
  Fundamental Physics, PHYSICS AT LEP}, Lekeitio, April 23-29, 1989, edited by
  M.A.-Ben{\'\i}tez and M. Cerrada, (World Scientific, Singapore, 1990), p.71.\\
   ALEPH Collaboration, D. Decamp {\it et al.}, {\sl Phys. Lett.}
  {\bf 244B} (1990), 541.\\
   DELPHI Collaboration, P. Abreu {\it et al.}, {\sl Phys. Lett.}
  {\bf 247B} (1990), 157.\\
   {\sl Proceedings of the Joint International Lepton-Photon Symposium
  \& Europhysics Conference on High Energy Physics}, Geneva, Switzerland,
  July 25-August 1, 1991, edited by S. Hegarty, K. Potter and E. Quercigh,
  (World Scientific, Singapore, 1992).
  \bibitem{tub} R. Barbieri and G. Giudice, {\sl Nucl. Phys.} {\bf B296}
  (1988), 75.\\
    T. Kon and M. Jimbo, in {\sl Proceedings of the First Workshop on
   Japan Linear Collider (JLC)}, KEK, October 24-25, 1989, edited by
   S. Kawabata, {\sl KEK Report} {\bf 90-2} (1990), p.280.\\
    M. Jimbo, in {\sl Proceedings of the Second Workshop on Japan Linear
   Collider (JLC)}, KEK, November 6-8, 1990, edited by S. Kawabata,
   {\sl KEK Proceedings} {\bf 91-10} (1991), p.185.
  \bibitem{col} {\sl Proceedings of the workshop on Physics at Future
   Accelerators}, La Thuile and CERN, January 1987, edited by J.H. Mulvey,
    CERN Report {\sl CERN} {\bf 87-07} (1987).\\
    C. Ahn {\it et al.}, {\sl SLAC-Report-}{\bf 329} (1988).\\
    {\sl Proceedings of the Third Workshop on Japan Linear Collider (JLC)},
    KEK, February 18-20, 1992, edited by A. Miyamoto, {\sl KEK Proceedings}
   {\bf 92-13} (1992).
  \bibitem{Rev} H. Bear {\it et al.}, preprint {\sl FSU-HEP-}{\bf 950401}
  ({\sl LBL-}{\bf 37016}, {\sl UH-}{\bf 511-822-95} and {\sl hep-ph/} {\bf
  9503479}) (1995), and References therein.
 \bibitem{tks} H. Tanaka, T. Kaneko and Y. Shimizu, {\sl Comput. Phys. Commun.}
  {\bf 64} (1991), 149.
  \bibitem{wmh} I. Watanabe, H. Murayama and K. Hagiwara, in {\sl Proceedings
  of the Third Workshop on Japan Linear Collider (JLC)}, KEK, February 18-20,
  1992, edited by A. Miyamoto, {\sl KEK Proceedings} {\bf 92-13} (1992), p.265.
  \bibitem{chan} H. Tanaka, {\sl Comput. Phys. Commun.} {\bf 58} (1990), 153.
  \bibitem{pre-grc} T. Kaneko, in {\sl New Computing Techniques in Physics
  Research}, edited by D. Perret-Gallix and W. Wojcik, ({\' E}dition du CNRS,
  Paris, 1990), p.555.\\
   T. Kaneko and H. Tanaka, in {\sl Proceedings of the Second Workshop
  on Japan Linear Collider (JLC)}, KEK, November 6-8, 1990, edited by
  S. Kawabata, {\sl KEK Proceedings} {\bf 91-10} (1991), p.250.\\
   T. Kaneko, in {\sl New Computing Techniques in Physics Research II},
  edited by D. Perret-Gallix, (World Scientific, Singapore, 1992), p.659.\\
   T. Ishikawa, T. Kaneko, K. Kato, S. Kawabata, Y. Shimizu and H. Tanaka
   (Minami-Tateya group), {\sl GRACE manual Version 1.0},
  {\sl KEK Report} {\bf 92-19} (1993), and References therein.
\bibitem{comp}
 E. Boos, M. Dubinin, V. Edneral, V. Ilyin, A. Kryukov, A. Pukov, V. Savrin,
 S. Shichanin and A. Taranov, in {\sl New Computing Techniques in Physics
 Research}, edited by D. Perret-Gallix and W. Wojcik, ({\' E}dition du CNRS,
 Paris, 1990), p.573.\\
 E. Boos, M. Dubinin, V. Edneral, V. Ilyin, A. Kryukov, A. Pukov, S. Shichanin,
 in {\sl New Computing Techniques in Physics Research II}, edited by
 D. Perret-Gallix, (World Scientific, Singapore, 1992), p.665.\\
 A. Pukov, in {\sl New Computing Techniques in Physics Research III}, edited by
 K.-H. Becks and D. Perret-Gallix, (World Scientific, Singapore, 1994), p.473.
 \bibitem{Den}
 J. K\"{u}blbeck, M. B\"{o}hm and A. Denner, {\sl Comput. Phys. Commun.}
 {\bf 60} (1990), 165.\\
 R. Mertig,  M. B\"{o}hm and A. Denner, {\sl Comput. Phys. Commun.}
 {\bf 64} (1991), 345.\\
 R. Mertig, in {\sl New Computing Techniques in Physics Research III} , edited
 by K.-H. Becks and D. Perret-Gallix, (World Scientific, Singapore, 1994),
 p.467.\\
 H. Eck and J. K\"{u}blbeck, {\it ibid.}, p.565.
 \bibitem{SL}
 T. Stelzer and W.F. Long, {\sl Comput. Phys. Commun.} {\bf 81}
  (1994), 357.
\bibitem{Den2}
 A. Denner, H. Eck, O. Hahn and J. K\"{u}blbeck, {\sl Phys. Lett.}
 {\bf B 291} (1992), 278.\\
 A. Denner, H. Eck, O. Hahn and J. K\"{u}blbeck, {\sl Nucl. Phys.}
 {\bf B 387} (1992), 467.
 \bibitem{bas} S. Kawabata, {\sl Comput. Phys. Commun.} {\bf 41} (1986), 127.\\
  S. Kawabata, {\sl Comput. Phys. Commun.} {\bf 88} (1995), 309.
  \bibitem{jt} M. Jimbo and H. Tanaka, {\sl Talk presented at JPS meeting},
  Fukuoka, March 28-31, 1994.
  \bibitem{jtkk} M. Jimbo, H. Tanaka, T. Kaneko, T. Kon and Minami-Tateya
  collaboration, in {\sl Physics of $e^+e^-$, $e^-\gamma$ and $\gamma\gamma$
  collisions at linear accelerators --- Proceedings of the INS Workshop}, INS,
  December 20-22, 1994, edited by Z. Hioki, {\it et al.}, {\sl INS-J-}{\bf 181}
  (1995), p.222.
  \bibitem{gm} Minami-Tateya collaboration, {\sl The document file for GRACE
  version~1.1},\\
  kek/minami/grace/grace.tar.Z at {\sl ftp.kek.jp} (130.87.34.28), (1994).
  \bibitem{grc-pp} T. Kaneko, in {\sl New Computing Techniques in Physics
  Research IV --- Proceedings of the Fourth International Workshop on
  Software Engineering, Artificial Intelligence and Expert Systems for High
  Energy and Nuclear Physics (AIHENP95)}, Pisa, Italy, April 3-8, 1995,
  edited by B. Denby and D. Perret-Gallix, (World Scientific, Singapore, 1995),
  p.313.
 \bibitem{Fdg} T. Kaneko, {\sl Comput. Phys. Commun.} {\bf 92} (1995), 127.
 \bibitem{sk} T. Ishikawa, S. Kawabata and Y. Kurihara, in {\sl Proceedings
  of the Fifth Workshop on Japan Linear Collider (JLC)}, Kawatabi, Miyagi,
  February 16-17, 1995, edited by Y. Kurihara, {\sl KEK Proceedings} {\bf 95-11}  (1995), p.92.
 \bibitem{grc2.0b} T. Ishikawa, S. Kawabata, Y. Kurihara and T. Kaneko, {\sl
 Brief Manual of Grace System Version 2.0 $\beta$} (1995), unpublished.
 \bibitem{YITP} M. Jimbo, T. Kon and Minami-Tateya collaboration, in {\sl
  Proceedings of the YITP Workshop on Particle Physics and its Future
  Perspective}, YITP, January 17-20, 1995, edited by K. Suehiro, {\sl 
  Soryushiron Kenkyu} {\bf 92} (1995), p.31.
  \bibitem{mj} M. Jimbo, in {\sl Proceedings of the Second Workshop on Japan
  Linear Collider (JLC)}, KEK, November 6-8, 1990, edited by S. Kawabata,
  {\sl KEK Proceedings} {\bf 91-10} (1991), p.185.
  \bibitem{Mor} T. Kaneko, H. Tanaka, M. Jimbo, T. Kon and Minami-Tateya
   collaboration, to appear in {\sl ELECTROWEAK INTERACTIONS AND UNIFIED
   THEORIES --- Proceedings of XXXth Rencontres de Moriond}, Les-Arc,
   France, March 11-18, 1995.\\
   See also {\sl http://jlcux1.kek.jp/subg/susy/lib/DOC/Grace/grace.html}.
  \bibitem{ai95} M. Jimbo, T. Kon, H. Tanaka, T. Kaneko and Minami-Tateya
  collaboration, in {\sl New Computing Techniques in Physics Research IV ---
  Proceedings of the Fourth International Workshop on Software Engineering,
  Artificial Intelligence and Expert Systems for High Energy and Nuclear
  Physics (AIHENP95)}, Pisa, Italy, April 3-8, 1995, edited by B. Denby and
  D. Perret-Gallix, (World Scientific, Singapore, 1995), p.149.
  \bibitem{prep} H. Tanaka, T. Kon, M. Jimbo, T. Kaneko and Minami-Tateya
  collaboration, in preparation.
  \bibitem{Appi} T. Kon, {\sl Talk presented at Workshop on Physics and
   Experiments with Linear Colliders}, Morioka-Appi, Iwate, Japan, September
   8-12, 1995.
  \bibitem{HESP} M. Jimbo, in {\sl Frontiers of High Energy Spin Physics ---
   Proceedings of the 10th International Symposium on High Energy Spin Physics
   (Yamada Conference XXXV)}, Nagoya, Japan, Novemver 9-14, 1992, edited by
   T. Hasegawa, N. Horikawa, A. Masaike and S. Sawada, (Universal Academy
   Press, 1993), p.657.\\
   M. Jimbo, {\sl Memoirs of Tokyo Management College} {\bf I} (1993), 101,
   and References therein.
  \bibitem{tk} T. Kon, {\sl Prog. Theor. Phys.} {\bf 79} (1988), 1006,
   and References therein.
  \bibitem{spu} M. Jimbo, T. Kon and T.Ochiai, Rikkyo University preprint
  {\sl RUP-}{\bf 87-1} (1987).\\
   M. Jimbo, {\sl Prog. Theor. Phys.} {\bf 79} (1988), 899, and References
   therein.
  \bibitem{epa} M. Jimbo and M. Katuya, {\sl Europhys. Lett.} {\bf 16}
   (1991), 243.\\
   M. Jimbo and M. Katuya, in {\sl Proceedings of the KEK Summer Institute on
   High Energy Phenomenology}, KEK, August 21-25, 1990, edited by K. Hikasa,
   {\sl KEK Proceedings} {\bf 91-8} (1991), p.84, and References therein.
  \bibitem{jlc5} M. Jimbo and Minami-Tateya collaboration, in {\sl Proceedings
  of the Fifth Workshop on Japan Linear Collider (JLC)}, Kawatabi, Miyagi,
  February 16-17, 1995, edited by Y. Kurihara, {\sl KEK Proceedings} {\bf 95-11}  (1995), p.98.
\end{thebibliography}
\end{document}